\documentclass[12pt,preprint]{aastex}
\usepackage{epsfig}
                                                                                
\def\vkm{km s$^{-1}$}

\def\degree{$^\circ$}
\def\arcs#1{$#1''$}
\def\arcsa#1#2{$#1^{\prime\prime}_{^\textrm{.}}#2$}

\def\solarmass{$M_\odot$}

\def\mJyb{mJy beam$^{-1}$}

\def\mJybk{mJy beam$^{-1}$ km s$^{-1}$}

\def\cms{cm$^{-2}$}
\def\micron{$\mu$m}
\def\VLSR{V_\textrm{\scriptsize LSR}}
\def\Vsys{V_\textrm{\scriptsize sys}}
\def\Voff{V_\textrm{\scriptsize off}}

\def\mH2{m_{\textrm{\scriptsize H}_2}}

\def\scnum#1#2{#1$\times10^{#2}$}
\def\H2{H$_2$}
\def\N2HP{N$_2$H$^+$}

\def\NH3{NH$_3$}

\def\SOTwo{SO$_2$}
\def\DtCO{D$_2$CO}

\def\HtCCO{H$_2$CCO}
\def\CHtDOH{CH$_2$DOH}
\def\CHtDCN{CH$_2$DCN}

\def\CHtOH{CH$_3$OH}
\def\tCHtOH{$^{13}$CH$_3$OH}
\def\CHtCHO{CH$_3$CHO}
\def\CHtOCHO{CH$_3$OCHO}
\def\CHtOCHO{HCOOCH$_3$}
\def\CHtCHtOH{CH$_3$CH$_2$OH}
\def\NHtCHO{NH$_2$CHO}
\def\CHtSH{CH$_3$SH}

\def\putfiga#1#2#3{\epsfig{scale=#1,angle=#2,figure=#3}}
\def\putfig#1#2#3{}
\def\leftblank#1{}
\def\bf#1{#1}

\begin{document}



\title{First Abundance Measurement of Organic Molecules in the Atmosphere of
HH 212 Protostellar Disk}


\author{Chin-Fei Lee\altaffilmark{1,2}, Claudio Codella\altaffilmark{3,4}, Zhi-Yun Li\altaffilmark{5},
and Sheng-Yuan Liu\altaffilmark{1}}

\altaffiltext{1}{Academia Sinica Institute of Astronomy and Astrophysics,
P.O. Box 23-141, Taipei 106, Taiwan; cflee@asiaa.sinica.edu.tw}
\altaffiltext{2}{Graduate Institute of Astronomy and Astrophysics, National Taiwan
   University, No.  1, Sec.  4, Roosevelt Road, Taipei 10617, Taiwan}
\altaffiltext{3}{INAF, Osservatorio Astrofisico di Arcetri, Largo E. Fermi 5,
50125 Firenze, Italy}
\altaffiltext{4}{Univ. Grenoble Alpes, CNRS, Institut de
Plan\'etologie et d'Astrophysique de Grenoble (IPAG), 38000 Grenoble, France}
\altaffiltext{5}{Astronomy Department, University of Virginia, Charlottesville, VA 22904, USA}

\begin{abstract}

HH 212 is one of the well-studied protostellar systems, showing the first
vertically resolved disk with a warm atmosphere around the central
protostar.  Here we report a detection of 9 organic molecules (including
newly detected ketene, formic acid, deuterated acetonitrile, methyl formate,
and ethanol) in the disk atmosphere, confirming that the disk atmosphere is,
for HH 212, the chemically rich component, identified before at a lower
resolution as a ``hot-corino''.  More importantly, we report the first
systematic survey and abundance measurement of organic molecules in the disk
atmosphere within $\sim$ 40 au of the central protostar.  The relative
abundances of these molecules are similar to those in the hot corinos around
other protostars and in Comet Lovejoy.  These molecules can be either (i)
originally formed on icy grains and then desorbed into gas phase or (ii)
quickly formed in the gas phase using simpler species ejected from the dust
mantles.  The abundances and spatial distributions of the molecules provide
strong constraints on models of their formation and transport in star
formation.  These molecules are expected to form even more complex organic
molecules needed for life and deeper observations are needed to find them.

\end{abstract}






\section{Introduction}

Accretion disks have been detected in very young protostellar systems,
feeding the central protostars.  With the advent of the Atacama Large
Millimeter/submillimeter Array (ALMA), we have started to resolve the disks
and study their physical processes in great detail.  HH 212 is one of
the well-studied protostellar systems, showing the first vertically resolved
disk with a warm atmosphere around the central protostar
\citep{Lee2017Disk,Lee2017COM}.  This warm atmosphere seems to be the hot
corino reported recently at a lower angular resolution
\citep{Codella2016,Codella2018}.  Hot corinos are the hot ($\gtrsim 100$ K)
and compact regions immediately around low-mass (sun-like) protostars
\citep{Ceccarelli2007}, and rich in organic molecules including complex
organic molecules \cite[COMs, refering to C-bearing species with six atoms
or more,][]{Herbst2009}.  By determining the connection of the disk
atmosphere with the hot corino in HH 212, we aim to determine the origin of
the hot corino and the related physical processes in the innermost region. 
In particular, the nearly edge-on orientation of the disk in this system
provides the best view of the atmosphere, allowing us to study the physical
properties of the disk atmosphere, and the formation of the organic
molecules and their role in producing the rich organic chemistry needed for
life.



HH 212 is a young Class 0 protostellar system deeply embedded in a compact
molecular cloud core in the L1630 cloud of Orion at a distance of $\sim$ 400
pc \citep{Kounkel2017}.  The central source is IRAS 05413-0104, with a
bolometric luminosity of $\sim$ 9 $L_\odot$ (updated for the new distance)
\citep{Zinnecker1992}.  The central protostar has a mass of $\sim$ 0.25
\solarmass{} \citep{Lee2017COM}.  It drives a powerful bipolar jet
\citep{Zinnecker1998,Lee2015}, which is recently found to be spinning
\citep{Lee2017Jet}.  A rotating disk must have formed around the protostar
in order to launch the jet.  Our previous ALMA observations towards the
center indeed showed a spatially resolved nearly edge-on dusty disk with a
radius of $\sim$ 60 au \citep{Lee2017Disk}.  In addition, we also detected a
warm atmosphere of the disk with a few organic molecules \citep{Lee2017COM},
suggesting that the warm disk atmosphere can be the hot corino reported
before at a lower resolution \citep{Codella2016}.  Recent observations at a
resolution of $\sim$ \arcsa{0}{15} (60 au) suggested that deuterated
water and \CHtCHO{} can also reside in the disk atmosphere
\citep{Codella2018}.  In this paper, we zoom in to the disk region at a
higher resolution of $\sim$ \arcsa{0}{03} (12 au) and higher sensitivity and
detect additional and more complex organic molecules characteristic of a hot
corino, with most of them detected for the first time in the disk
atmosphere.  Our observations confirm that the hot corino in HH 212 is
indeed the warm disk atmosphere.  We will discuss the formation of the organic
molecules by comparing their abundances to those in hot corinos around other
low-mass protostars.

\section{Observations}\label{sec:obs}

The HH 212 protostellar system was observed with ALMA in Band 7 at $\sim$
341.5 GHz in Cycle 5.  Project ID was 2017.1.00044.S.  Three observations
were executed on the same day on 2017 November 27 with 47 antennas, with
each observation having an integration time of $\sim$ 32.7 min on the
target.  Thus, the total time on the target was $\sim$ 98 minutes.  The
projected baselines were $\sim$ 60-8500 m.  The correlator was set up to
have 4 spectral windows, with 3 in continuum mode having a bandwidth of 2
GHz, and one in spectral mode having a bandwidth of 1.875 GHz.  In this
paper, we only report the results obtained with the spectral mode that
covers the rest frequency from $\sim$ 345.635 to 347.510 GHz.  The spectral
resolution was 1.953 MHz per channel, resulting in a velocity resolution of
$\sim$ 1.69 \vkm{} per channel at 346.5 GHz.  The primary beam was
\arcsa{17}{27}.  A single pointing was used to map the system within $\sim$
\arcs{1} of the central source.  The maximum recoverable size scale was
$\sim$ \arcsa{0}{4}, enough to map the disk atmosphere without any
significant missing flux.

%

The data were calibrated with the CASA package, with quasar J0510+1800 as a
passband and flux calibrator, and quasar J0541-0211 (a flux of 0.137 Jy) as
a gain calibrator.  We used a robust factor of 0.5 for the visibility
weighting to generate the channel maps with a synthesized beam of
\arcsa{0}{036}$\times$\arcsa{0}{03} at a position angle of $\sim$
$-$78\degree{}.  We used the line-free channels to generate a continuum map
centered at 356.5 GHz.  The channel maps of the molecular lines were
generated after continuum subtraction.  The noise levels are $\sim$ 0.121
\mJyb{} (or $\sim$ 1.13 K) in the continuum map and 0.80 \mJyb{} (or $\sim$
7.45 K) in the channel maps.  The velocities in the channel maps are LSR
velocities.

\section{Results}

In HH 212, the jet has an axis with a position angle of $\sim$ 23\degree{} and an
inclination angle of $\sim$ 4\degree{} to the plane of the sky, with its
northern component tilted toward us \citep{Claussen1998}.  The disk is
nearly edge-on and exactly perpendicular to the jet axis
\citep{Lee2017Disk}.  The systemic velocity is $\Vsys= 1.7\pm0.1$ \vkm{} LSR
\citep{Lee2014}.  In order to facilitate our presentations, we define an
offset velocity $\Voff = \VLSR - \Vsys$ and rotate our maps by 23\degree{}
clockwise to align the jet axis in the vertical direction.





\begin{figure} [!hbp]
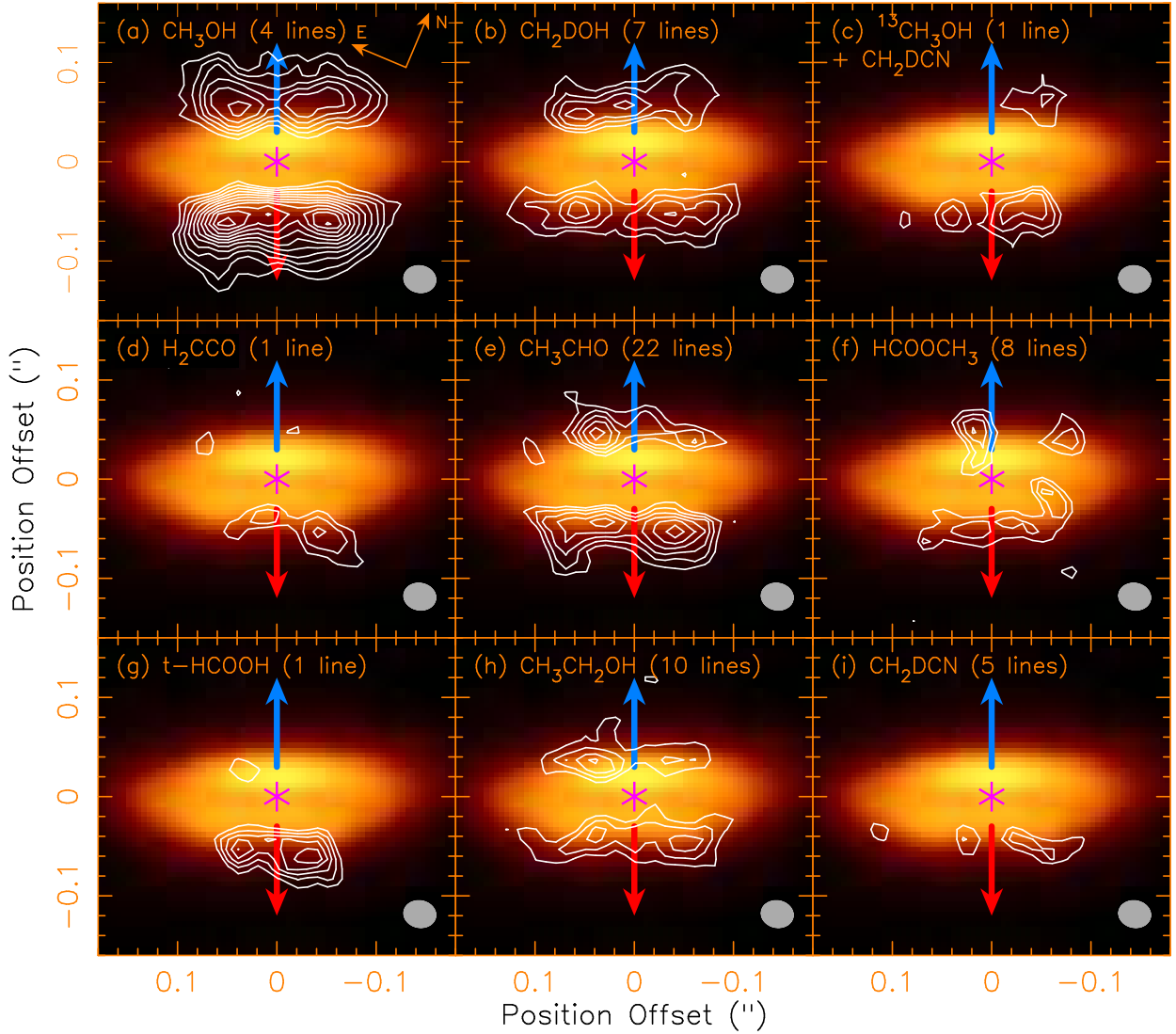

\centering
\putfiga{0.85}{270}{f1.eps} 
\figcaption[]
{Total emission line intensity maps of nine organic molecules integrated 
over velocity within 
$\sim$ 4 \vkm{} of the systemic velocity,
plotted on top of the continuum map (color image) of the dusty disk.
As discussed in the text, the maps of the molecular emissions are obtained by
stacking a number of lines (as indicated in the parenthesis) in
different transitions with a range of upper energy levels for better
detections. The asterisk marks the possible position of the central protostar.
The blue and red arrows indicate the axes of the blueshifted and redshifted jet components,
respectively. The contours start at 3 $\sigma$ with a step of 1 $\sigma$.
The 1-$\sigma$ noise levels in panels (a) to (i) 
are 1.96, 1.64, 4.06, 3.91, 0.89, 1.10, 4.23, 1.19, and 1.72 \mJybk{}.
\label{fig:contCOMs}}
\end{figure}

Figure \ref{fig:contCOMs} shows the emission line intensity maps (contours)
of nine organic molecules, including \CHtOH{} (methanol, 4 lines), \CHtDOH{}
(deuterated methanol, 7 lines), \tCHtOH{} ($^{13}$C isotopologue of
methanol, 1 line, contaminated by a much weaker \CHtDCN{} line), \HtCCO{}
(ketene, 1 line), \CHtCHO{} (acetaldehyde, 22 lines), \CHtOCHO{} (methyl
formate, 8 lines), t-HCOOH (formic acid in {\it trans} state, 1 line),
\CHtCHtOH{} (ethanol, 10 lines), and \CHtDCN{} (deuterated acetonitrile, 5
lines), on the continuum map (color image) of the disk at $\lambda \sim$ 865
\micron{} (or equivalently $\sim$ 347 GHz), obtained from our observations. 
As seen before at a similar wavelength ($\lambda \sim$ 850 \micron{}) in
\citet{Lee2017Disk}, the continuum map shows a ``hamburger-shaped'' dusty
disk with an equatorial dark lane sandwiched between two brighter features
on the top and bottom.  As discussed in that paper, the presence of the
equatorial dark lane is due to relatively low temperature and high optical
depth near the disk midplane.  As discussed below, the maps of the molecular
emissions are obtained by stacking a number of lines (as indicated above in
the parenthesis) in different transitions with a range of upper energy
levels for better detections.  Molecular line emissions are only detected in
the upper (above the midplane) and lower disk atmosphere, with the emission
brighter in the lower disk atmosphere below the midplane.  No molecular
emission is detected toward the disk midplane, either because the emission
is lost in the optically thick dust continuum emission or because of a lack
of emission of these molecules there.  All the molecular emissions are
detected within the centrifugal barrier (which has a radius of $\sim$
\arcsa{0}{11} or 44 au) of the central protostar.  For \CHtOH{}, \CHtDOH{},
\CHtCHO{}, and \CHtCHtOH{}, their emissions are clearly detected in the
upper and lower atmosphere with a roughly similar distribution, suggesting
that they are chemically related.  In addition, the emission moves closer to
the disk midplane from \CHtOH{}, \CHtDOH{}, \CHtCHO{}, to \CHtCHtOH{},
suggesting that the emission of less abundant molecule (see next section for
their abundances) traces deeper into the disk atmosphere, probably due to an
optical depth effect.  However, it could also be due to a chemical
stratification in the vertical direction.  \CHtDOH{} and \CHtCHtOH{}
emission in the lower atmosphere in the outer edge show a structure curving
back towards the midplane, following the lower boundary of the dusty disk
emission, likely outlining a physical boundary for the dusty disk.





Figure \ref{fig:pv_atms} shows the position-frequency (PF) diagrams obtained
with a cut along the lower atmosphere, where the emission is brighter, in
order to identify the detections of various molecular lines.  The PF
diagrams for the upper atmosphere show similar structures but fainter and
are thus not shown here.  As can be seen, many lines are detected, with
their emission detected within $\sim$ 5 MHz of their rest frequencies (or
$\sim$ 4 \vkm{} within the systemic velocity) marked by the vertical lines,
with one color for each organic molecular species at different transitions. 
For each line, the diagrams show a roughly linear PF structure with the
redshifted emission in the southeast (positive $y$) and blueshifted emission
in the northwest (negative $y$), as seen before in the corresponding
position-velocity diagrams for the \CHtOH{} and \CHtDOH{} lines
\citep{Lee2017COM}.  As discussed in that paper, this linear PF structure
likely arises from a warm rotating ring of the disk atmosphere near the
centrifugal barrier.  With this feature, we identify lines from the nine
organic molecules mentioned above, and a few other simple molecules (e.g.,
SO, \SOTwo{} and its isotopologue SO$^{18}$O).  Other lines from CO and SiO
trace mainly the outflow and jet, and thus do not show such a clear linear PF
structure.


Table \ref{tab:lines} lists the properties of the organic molecular lines. 
For each molecular species, we stacked the line intensity maps in different
transitions, excluding those tentatively detected (marked with T) and
blended (marked with B), producing the mean line intensity map shown in
Figure \ref{fig:contCOMs}.

\subsection{Physical condition in the disk atmosphere}

Here we derive the mean excitation temperature and column density in the
disk atmosphere by fitting the rotation diagram of the molecular lines. 
This diagram plots the column density per statistical weight in the upper
energy state in the optically thin limit, $N_u^\textrm{\scriptsize
thin}/g_u$, versus the upper energy level $E_u$ of the lines.  Here
$N_u^\textrm{\scriptsize thin}=(8\pi k\nu^2/hc^3 A_{ul}) W$, where the
integrated line intensity $W =\int T_B dv$ with $T_B$ being the brightness
temperature.



The emission in the lower disk atmosphere is brighter and is thus used to
better derive the mean excitation temperature and column density in the disk
atmosphere.  Table \ref{tab:lines} lists the integrated line intensities in
the lower disk atmosphere measured (with a cutoff of 2$\sigma$) for the
reasonably isolated lines detected with more than $3\sigma$.  They are the
mean values averaged over a rectangular region that covers most of the
emission in the lower atmosphere.  Figure \ref{fig:pop} shows the resulting
diagrams for the six molecules detected with multiple lines.  The blended
lines are excluded.  With the \CHtOH{} lines, we derived an excitation
temperature of $\sim$ $160\pm45$ K.  With the lines from its deuterated
species \CHtDOH{}, we derived a similar temperature of $\sim$ $148\pm19$ K. 
For a check, we also obtained similar temperatures of $\sim$ $138\pm32$ K
and $\sim$ $168\pm24$ K for the upper atmosphere from the \CHtOH{} lines and
\CHtDOH{} lines, respectively.  The resulting column densities of \CHtOH{}
and \CHtDOH{} are listed in Table \ref{tab:colabun}.   As discussed
later, since the \CHtOH{} lines are likely optically thick, the \CHtOH{}
column derived here is only a lower limit and a more accurate value will be
derived using the \tCHtOH{} column density below.




From \CHtOH{} and \CHtDOH{}, we obtain a mean excitation temperature of
$\sim$ $150\pm50$ K in the lower disk atmosphere, similar to that found
before in \citet{Lee2017COM}.  This mean temperature is also consistent
with that derived towards the disk at a lower angular resolution in
\citet{Bianchi2017}.  Assuming this mean temperature for the disk
atmosphere, we can estimate the column densities of other molecules with
weaker lines.  For those molecules detected with multiple lines, such as
\CHtCHO{}, \CHtOCHO{}, \CHtCHtOH{}, and \CHtDCN{}, we can obtain their
column densities by fitting their rotation diagrams, as shown in Figures
\ref{fig:pop}(c)-(f).  Notice that the column density of \CHtCHO{} is
estimated here to be  \scnum{(1.5$\pm$0.6)}{16} \cms{}, with the lower
limit consistent with that estimated previously at a lower resolution with
an excitation temperature of $\sim$ 78$\pm$14 K in \citet{Codella2018}. 
For those detected with a single line, such as \tCHtOH{}, \HtCCO{}, and
t-HCOOH, we derived their column densities from their measured integrated
line intensity.  For \tCHtOH{}, the only detected line is contaminated by a
weaker line of \CHtDCN{} and thus its column density is estimated after
removing the expected intensity of the \CHtDCN{} line.  The expected
intensity of the \CHtDCN{} line is assumed to be given by the mean intensity
of other \CHtDCN{} lines with similar E$_u$ and $\log A_{ul}$, and it is
estimated to be $\sim$ 21 K \vkm{}, or $\sim$ 18\% of the total intensity
there.   Also, since the \tCHtOH{} line is optically thinner than the
\CHtOH{} lines, we can improve the \CHtOH{} column density by multiplying
the \tCHtOH{} column density by a $^{12}$C/$^{13}$C ratio of 50 as obtained
in the Orion Complex \citep{Kahane2018}.  As can be seen from Table
\ref{tab:colabun}, the \CHtOH{} column density derived this way is about
twice that derived from the rotation diagram, suggesting that the \CHtOH{}
lines are indeed optically thick. Also shown in the table are the
column densities of \NHtCHO{}, \DtCO{}, and \CHtSH{} measured in
\citet{Lee2017COM}, adjusted with the mean excitation temperature here.






Based on our disk model that reproduced the dust continuum emission at
a similar wavelength of $\sim$ 850 \micron{} \citep{Lee2017Disk}, the dust
continuum emission in the disk atmosphere has an optical depth $\tau
\lesssim 1$.  Thus, the derived column densities of the organic molecules
could be underestimated by a factor of $e$. In that disk model, the
disk atmosphere has a mean H$_2$ column density of $\sim$ \scnum{3.8}{24}
\cms{}.  Hence, the abundances of the molecules, as listed in Table
\ref{tab:colabun}, can be obtained by dividing the column density of the
molecules by this mean H$_2$ column density.  The abundances here can be
uncertain by a factor a few because the mean H$_2$ column density depends
on dust opacity, which can be uncertain by a factor a few.

\section{Discussion}

\subsection{Corino-like Disk Atmosphere}



Hot corinos with a temperature $\gtrsim$ 100 K have been detected around
low-mass protostars and they are rich in organic molecules (including
complex organic molecules).  A hot corino has also been recently reported in
HH 212 in the inner 100 au with a detection of \CHtCHO{} and deuterated
water HDO \citep{Codella2016,Codella2018}.  Now at higher resolution, we
find \CHtCHO{} to reside in the disk atmosphere.  Moreover, we also detect 8
other organic molecules (\HtCCO{}, t-HCOOH, \CHtOH{}, \CHtDOH{}, \tCHtOH{},
\CHtDCN{}, \CHtOCHO{}, and \CHtCHtOH{}) in the disk atmosphere, with an
excitation temperature of $\sim$ 150$\pm$50 K.  Adding 3 other organic molecules
(\DtCO{}, \CHtSH{} and \NHtCHO{}) from \citet{Lee2017COM}, we have 12
organic molecules, with 9 of them being complex organic molecules, detected
in the disk atmosphere.  These results indicate that the hot corino in HH
212 is actually a warm atmosphere of the disk.




\subsection{Formation of Organic Molecules}

\begin{figure} [!hbp]
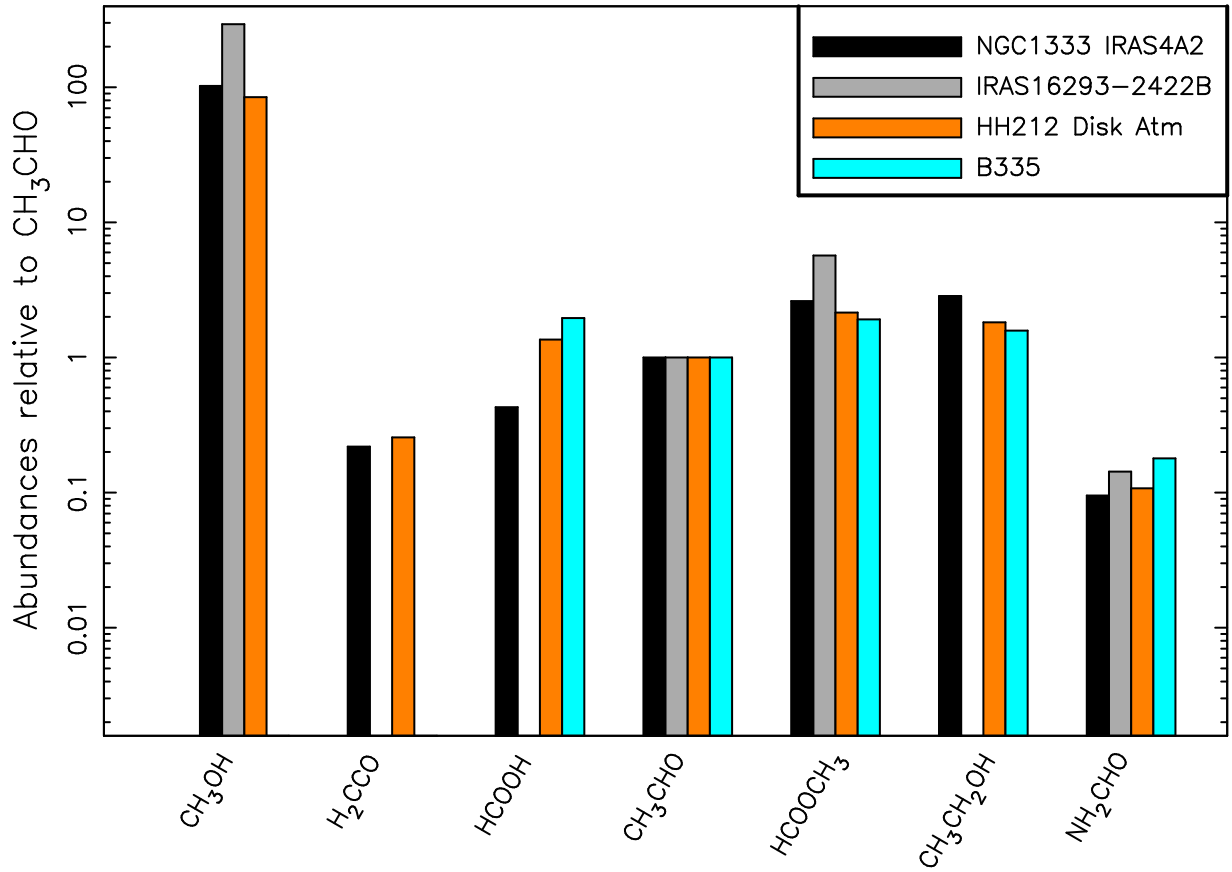

\centering
\putfiga{0.7}{270}{f4.eps} 
\figcaption[]
{A comparison for the abundances of organic molecules in the HH 212 disk atmosphere 
with those in the hot corinos around other low-mass protostars.
{\bf As shown in Table \ref{tab:colabun}, our measurements have an uncertainty of about 50\%.}
\label{fig:abunratio}}
\end{figure}

We can study the formation of the organic molecules in the disk atmosphere
by comparing their abundances to those estimated in the hot corinos around
other Class 0 low-mass protostars.  The abundance of \CHtCHO{}, which is
commonly observed in hot corinos, is estimated here to be \scnum{3.9}{-9}
in HH 212, similar to those found in other hot corinos, e.g.,
5.8$\times10^{-9}$ in IRAS 16293-2422B \citep{Jorgensen2016},
2.4$\times10^{-9}$ in B335 \citep{Imai2016}, and 4.2$\times10^{-9}$ in NGC
1333 IRAS4A2 \citep{Lopez2017}.  To facilitate comparison, we calculate and
compare the abundances of the organic molecules relative to \CHtCHO{}. 
Figure \ref{fig:abunratio} shows the comparison of the relative abundances
of 7 organic molecules (excluding deuterated species and $^{13}$C
isotopologue) to those in IRAS 16293-2422B \citep{Jorgensen2016}, B335
\citep{Imai2016}, and NGC 1333 IRAS4A2 \citep{Lopez2017}.  In this
comparison,  \CHtSH{} is excluded because of no
reliable measurement of this molecule in other hot corinos to compare with. 
Interestingly, the relative abundances of most molecules here in HH 212 are
similar to those in other hot corinos to within a factor of few, suggesting
that the formation of these molecules in the disk atmosphere could be
similar to that in other hot corinos.  Notice that the abundance of \CHtSH{}
here is also consistent with that predicted in the hot corino, which can be
as high as 5$\times10^{-8}$ \citep{Majumdar2016}.  Moreover, the abundances
here are also similar to those seen in the Class I corino SVS13-A
\citep{Bianchi2019} and to those seen in Comet Lovejoy, which shows similar
abundances to IRAS 16293-2422B \citep{Biver2015}.


As discussed in \citet{Lee2017COM}, the high degree of deuteration in
methanol {\bf (with [\CHtDOH]/[\CHtOH] $\sim 0.12$)}
and the detection of doubly deuterated formaldehyde suggest that the
methanol and formaldehyde in the disk atmosphere are originally formed
on icy grains and later desorbed (evaporated) into gas phase due to the
heating possibly by low-velocity accretion shock near the centrifugal
barrier or the radiation of the protostar.  This heating can also be
produced by an interaction with a wind/outflow from an inner region
\citep{Lee2018Dwind}.  The temperature of the disk atmosphere where the
organic molecules are detected is similar to the freeze-out temperature of
water, which is $\sim$ 150 K \citep{Oberg2011}.  Thus, the detection of
deuterated water in the disk atmosphere \citep{Codella2018} supports that
the two organic molecules are evaporated in the disk atmosphere
where it is warm enough to release the water.  However, how the other
organic molecules are formed is still an open question and not necessarily
on the grain surfaces.





In case the organic molecules are formed on the icy grains, they can be
formed on the icy grains in the disk, as suggested in the TW Hya
protoplanetary disk for the methanol \citep{Walsh2016}, either in situ on
the surface or first in the midplane and then brought to the surface by
turbulence \citep{Furuya2014}.  Like the hot corinos, it is also possible
that the molecules already formed on the icy grains in the prestellar core,
and then brought in to the disk surface.  CO is frozen onto to the grains at
the temperature of $\sim$ 20 K \citep{Oberg2011}.  {\bf One possible
scenario (although to be proven) is that, in the regions such as prestellar
cores and probably disk midplane where the temperature is below 20 K,
CO-rich ices on dust grains can undergo addition reactions with H (and D),
O, C, and N atoms accreted from the gas, producing a rich organic chemistry
on the grains, as proposed in \citet{Charnley2008}.  }


Having said that, some of the organic molecules could also be quickly formed
in the gas phase using simpler species released from grains.  This could be
the case for formamide, which was found to be mainly formed in the gas phase
in a young shocked region such as L1157-B1 \citep{Codella2017}.   In
addition, acetaldehyde can also be formed in gas phase, as suggested in
\citet{Charnley2004}.

These organic molecules are of great importance for forming even more
complex organic molecules such as amino acids and amino sugars, which are
the building blocks of life.  Our observations clearly show that they have
formed in a disk or been brought in to a disk in the earliest phase of star
formation and may play a crucial role in producing the rich organic
chemistry needed for life.  It is also tempting to estimate the alcohol
degree in the disk atmosphere.  According to \citet{Codella2018}, the column
density of deuterated water is $\lesssim 3\times10^{17}$ \cms{}.  Assuming
[H/D] $\sim$ [\CHtOH/\CHtDOH] $\sim$ 8.1, then the column density of the
water is $\lesssim 2.4\times10^{18}$ \cms{}.  With the derived column
density of ethanol in Table \ref{tab:colabun}, and the molecular mass of
ethanol of 46 and water of 18, the alcohol degree by mass is estimated to be
$\gtrsim$ 2.8\%, and thus could be similar to that of a regular beer.

\subsection{Possible Disk Wind?}

Methanol and acetaldehyde have been argued to trace a disk wind in HH 212
\citep{Leurini2016,Codella2018}.  Now in observations at higher
resolution, they seem to trace mainly a ring of disk atmosphere near the
centrifugal barrier, which could in principle be heated by a weak
(accretion) shock produced by the rapid decrease of the infall velocity near
the centrifugal barrier.  However, since their maps also show small
extensions extending out from the disk (see Figure \ref{fig:contCOMs}), they
may also trace a wind coming out from the disk surface.  Nonetheless, since
these extensions appear to be surrounding the SO outflow shell detected
further in \citet{Lee2018Dwind}, they could also be the disk atmosphere
pushed away by the SO outflow shell.  Detailed kinematic study of these
extensions with the SO outflow shell are needed to check these scenarios.

\section{Conclusions}


The nearly edge-on orientation of the disk in HH 212 provides the best view
of the disk atmosphere.  Here we have detected 9 organic molecules in the
disk atmosphere.  These molecules are characteristic of a hot corino and
found here to be in the disk atmosphere, confirming that the corino here is
a warm disk atmosphere.  Adding 3 other organic molecules from our previous
study, we have detected 12 organic molecules, with 9 of them being complex
organic molecules, in the disk atmosphere within $\sim$ 40 au of the central
protostar.  These molecules seem to arise mainly from a ring of disk
atmosphere near the centrifugal barrier.  Some of them may also trace a wind
coming out from the disk surface.

The relative abundances of the organic molecules in the HH 212 disk
atmosphere are similar to those in hot corinos around other low-mass
protostars and even to those in Comet Lovely.  It would be interesting to
determine whether the hot corinos around other low-mass protostars are also
located in their disk atmospheres or not, perhaps through higher resolution
ALMA observations.  In addition, the formation mechanism of the organic
molecules can also be similar to that in those corinos.  The organic
molecules can originally formed on icy grains, either in the disk or in the
prestellar core and then brought in to the disk, and then desorbed
(evaporated) into the gas phase.  They can also be quickly formed in the gas
phase using simpler species ejected from the dust mantles.

\acknowledgements


{\bf We thank the anonymous referee for constructive comments.}
This paper makes use of the following ALMA data:
ADS/JAO.ALMA\#2017.1.00044.S.  ALMA is a partnership of ESO (representing
its member states), NSF (USA) and NINS (Japan), together with NRC (Canada),
NSC and ASIAA (Taiwan), and KASI (Republic of Korea), in cooperation with
the Republic of Chile.  The Joint ALMA Observatory is operated by ESO,
AUI/NRAO and NAOJ.  C.-F.L.  acknowledges grants from the Ministry of
Science and Technology of Taiwan (MoST 104-2119-M-001-015-MY3 and
107-2119-M-001 -040 -MY3) and the Academia Sinica (Career Development Award
and Investigator Award).  CC acknowledges the project PRIN-INAF 2016 The
Cradle of Life - GENESIS-SKA (General Conditions in Early Planetary Systems
for the rise of life with SKA).  ZYL is supported in part by NSF grant
AST-1815784 and ASTR-1716259 and NASA grant 80NSSC18K1095 and NNX14AB38G.

\def\nat{Natur}

\begin{figure} [!hbp]
\centering
\putfiga{0.8}{0}{f2.eps} 
\figcaption[]
{Position-frequency diagrams cut along the lower atmosphere of the disk,
in order to show the
detections of various molecular lines. The contours start from 2 $\sigma$ with a step of 1
$\sigma$ of 4 K.  The frequency has been corrected for the systemic velocity of
1.7 \vkm{}.  The vertical lines mark the rest frequencies of the lines in
different transitions, with one color for one organic molecular species.
The lines from simple molecules are marked with the black vertical lines.
The physical properties of the lines are listed in Table \ref{tab:lines}.
\label{fig:pv_atms}}
\end{figure}


\begin{figure} [!hbp]
\centering
\putfiga{0.7}{270}{f3.eps} 
\figcaption[]
{Rotation diagrams for molecular lines of the six organic molecules 
including \CHtOH{}, \CHtDOH{}, \CHtCHO{}, \CHtOCHO, \CHtCHtOH, and \CHtDCN{}.
The diagrams are derived from the line intensities in the lower disk atmosphere listed
in Table \ref{tab:lines}.
The error bars show the uncertainty in our measurements, which are assumed 
to be 40\% of the data values.
The solid line is a linear fit to the data. In panels (c)-(f), the temperature is fixed
at 150$\pm$50 K for the fitting in order to derive the column densities from the weak lines.
\label{fig:pop}}
\end{figure}

\clearpage

%

\begin{deluxetable}{llcccc}
\tablecolumns{8}
\tabletypesize{\normalsize}
\tablecaption{Line Properties from Splatalogue
 \label{tab:lines}}
\tablewidth{0pt}
\tablehead{
\colhead{Transition} & Frequency & 
    log($A_{ul}$) &     $E_{u}$ & $W^a$ & Line \\
\colhead{QNs}     & (MHz)  & 
    (s$^{-1}$) & (K) & (K \vkm{}) & List
}  
\startdata
t-HCOOH 15( 2,13)-14( 2,12)   &	346718.85 & -3.331 & 144.457 & 147 & CDMS \\
\\
\HtCCO{} 17( 1,16)-16( 1,15)  &	346600.45 & -3.325 & 162.789 & 79 & JPL \\
\\
\CHtOH{} 16(1)- - 15(2)-     & 345903.91 & -4.044 & 332.653	& 234 & JPL \\
\CHtOH{} 18(-3) - 17(-4), E2 & 345919.26 & -4.136 & 459.435	& 200 & JPL \\
\CHtOH{} 5(4)- - 6(3)-	     & 346202.71 & -4.662 & 115.162	& 129$^m$ & JPL \\
\CHtOH{} 5(4)+ - 6(3)+	     & 346204.27 & -4.662 & 115.162	& 129$^m$ & JPL \\
\\
\tCHtOH{} 14( 1,13)- 14( 0,14) - + & 347188.28 & -3.360 & 254.251 & 96$^\dagger$ & CDMS\\
\\
\CHtDOH{} 3(2,1) - 2(1,2), e1 	  & 345718.71 & -4.373 &  39.434 & 51 & JPL \\  
\CHtDOH{} 19(1,19) - 18(2,17), e1 & 345820.79 & -4.531 & 418.032 & B & JPL \\  
\CHtDOH{} 22(4,19) - 22(3,19), e1 & 345850.48 & -3.888 & 613.625 & 16$^m$ & JPL \\ 
\CHtDOH{} 3(1,2) - 3(0,3), e1	  & 346256.50 & -3.699 &  29.488 & 152 & JPL \\    
\CHtDOH{} 22(4,18) - 22(3,20), e1 & 346281.30 & -3.874 & 613.621 & 16$^m$ & JPL \\ 
\CHtDOH{} 21(4,18) - 21(3,18), e1 & 346419.06 & -3.886 & 566.562 & 13 & JPL \\     
\CHtDOH{} 21(4,17) - 21(3,19), e1 & 346783.70 & -3.876 & 566.559 & B & JPL \\ 
\CHtDOH{} 20(4,17) - 20(3,17), e1 & 346923.75 & -3.886 & 521.634 & 85 & JPL \\     
\CHtDOH{} 20(4,16) - 20(3,18), e1 & 347222.99 & -3.878 & 521.632 & B & JPL \\ 
\CHtDOH{} 19(4,16) - 19(3,16), e1 & 347371.16 & -3.887 & 478.841 & 18 & JPL \\     
\\
\CHtDCN{} 20( 1,20)-19( 1,19) &	345685.36 & -2.444 & 179.628 & 12 & JPL \\ 
\CHtDCN{} 20( 0,20)-19( 0,19) &	347043.43 & -2.438 & 174.955 & 27 & JPL \\ 
\CHtDCN{} 20( 4,17)-19( 4,16) &	347166.47 & -2.455 & 261.246 & 13$^m$ & JPL \\ 
\CHtDCN{} 20( 4,16)-19( 4,15) &	347166.48 & -2.455 & 261.246 & 13$^m$ & JPL \\ 
\CHtDCN{} 20( 2,19)-19( 2,18) &	347188.29 & -2.441 & 196.562 & B & JPL \\ 
\CHtDCN{} 20( 3,18)-19( 3,17) &	347216.91 & -2.447 & 223.527 & B & JPL \\  
\CHtDCN{} 20( 3,17)-19( 3,16) &	347219.39 & -2.447 & 223.527 & B & JPL \\  
\CHtDCN{} 20( 2,18)-19( 2,17) &	347388.21 & -2.441 & 196.615 & 24 & JPL \\ 
\CHtCHO{} vt=1, 18(2,17) - 17(2,16), E      & 346065.34 & -2.838 & 371.350 & 90 & JPL \\ 
\CHtCHO{} v=0, 18(11, 7) - 17(11, 6), E     & 346697.59 & -3.028 & 430.543 & T & JPL \\
\CHtCHO{} v=0, 18(10, 8) - 17(10, 7), E     & 346742.00 & -2.986 & 383.352 & 16 & JPL \\ 
\CHtCHO{} v=0, 18(11, 8) - 17(11, 7), E     & 346754.52 & -3.028 & 430.451 & 12$^m$ & JPL \\ 
\CHtCHO{} v=0, 18(11, 8) - 17(11, 7), E     & 346755.92 & -3.028 & 430.474 & 12$^m$ & JPL \\ 
\CHtCHO{} v=0, 18(11, 7) - 17(11, 6), E     & 346755.92 & -3.028 & 430.474 & 12$^m$ & JPL \\ 
\CHtCHO{} v=0, 18(10, 9) - 17(10, 8), E     & 346763.91 & -2.985 & 383.297 & 30 & JPL \\ 
\CHtCHO{} v=0, 18(9, 9) - 17(9, 8), E 	    & 346787.03 & -2.950 & 340.618 & B & JPL \\ 
\CHtCHO{} v=0, 18(10, 8) - 17(10, 7), E     & 346805.46 & -2.985 & 383.258 & S & JPL \\ 
\CHtCHO{} v=0, 18(10, 9) - 17(10, 8), E     & 346805.46 & -2.985 & 383.258 & S & JPL \\ 
\CHtCHO{} v=0, 18(9, 10) - 17(9, 9), E      & 346807.99 & -2.950 & 340.612 & S & JPL \\ 
\CHtCHO{} v=0, 18(8,10) - 17(8, 9), E 	    & 346839.03 & -2.920 & 302.406 & 23 & JPL \\ 
\CHtCHO{} v=0, 18(9,10) - 17(9, 9), E 	    & 346849.06 & -2.950 & 340.535 & 30$^m$ & JPL \\ 
\CHtCHO{} v=0, 18(9, 9) - 17(9, 8), E  	    & 346849.06 & -2.950 & 340.535 & 30$^m$ & JPL \\ 
\CHtCHO{} v=0, 18(8,11) - 17(8,10), E 	    & 346892.18 & -2.920 & 302.343 & S & JPL \\ 
\CHtCHO{} v=0, 18(8,11) - 17(8,10), E	    & 346893.81 & -2.920 & 302.316 & S & JPL \\ 
\CHtCHO{} v=0, 18(8,10) - 17(8, 9), E 	    & 346893.81 & -2.920 & 302.316 & S & JPL \\ 
\CHtCHO{} v=0, 18(7,11) - 17(7,10), E	    & 346934.22 & -2.896 & 268.661 & B & JPL \\ 
\CHtCHO{} v=0, 18(7,12) - 17(7,11), E	    & 346957.55 & -2.896 & 268.606 & B & JPL \\ 
\CHtCHO{} v=0, 18(7,11) - 17(7,10), E 	    & 346957.55 & -2.896 & 268.606 & B & JPL \\ 
\CHtCHO{} v=0, 18(7,12) - 17(7,11), E	    & 346995.53 & -2.896 & 268.572 & 24 & JPL \\ 
\CHtCHO{} v=0, 18(6,13) - 17(6,12), E	    & 347071.54 & -2.875 & 239.399 & 33$^m$ & JPL \\ 
\CHtCHO{} v=0, 18(6,12) - 17(6,11), E	    & 347071.68 & -2.875 & 239.399 & 33$^m$ & JPL \\ 
\CHtCHO{} v=0, 18(6,12) - 17(6,11), E 	    & 347090.40 & -2.875 & 239.397 & 33$^m$ & JPL \\ 
\CHtCHO{} v=0, 18(6,13) - 17(6,12), E	    & 347132.68 & -2.875 & 239.321 & 33 & JPL \\ 
\CHtCHO{} vt=1, 18(4,14) - 17(4,13), E 	    & 347182.41 & -2.845 & 400.378 & B & JPL \\ 
\CHtCHO{} vt=1, 18(5,13) - 17(5,12), E 	    & 347216.79 & -2.859 & 420.440 & B & JPL \\ 
\CHtCHO{} vt=1, 18(5,14) - 17(5,13), E	    & 347251.82 & -2.858 & 419.672 & T & JPL \\
\CHtCHO{} v=0, 18(5,14) - 17(5,13), E 	    & 347288.26 & -2.858 & 214.697 & 53 & JPL \\ 
\CHtCHO{} v=0, 18(5,13) - 17(5,12), E 	    & 347294.87 & -2.858 & 214.698 & 115 & JPL \\ 
\CHtCHO{} v=0, 18(5,13) - 17(5,12), E	    & 347345.71 & -2.858 & 214.640 & B & JPL \\ 
\CHtCHO{} v=0, 18(5,14) - 17(5,13), E 	    & 347349.27 & -2.858 & 214.611 & B & JPL \\ 
\\
\CHtOCHO{} v=0 28(12,16)-27(12,15) E & 345974.66 & -3.287 & 335.433 & 11$^m$ & JPL \\ 
\CHtOCHO{} v=0 28(12,17)-27(12,16) A & 345985.38 & -3.287 & 335.434 & 11$^m$ & JPL \\ 
\CHtOCHO{} v=0 28(12,16)-27(12,15) A & 345985.38 & -3.287 & 335.434 & 11$^m$ & JPL \\ 
\CHtOCHO{} v=0 28(12,17)-27(12,16) E & 346001.61 & -3.287 & 335.430 & 11$^m$ & JPL \\ 
\CHtOCHO{} v=0 28(11,17)-27(11,16) E & 346659.86 & -3.269 & 320.394 & B & JPL \\ 
\CHtOCHO{} v=0 28(11,18)-27(11,17) A & 346674.98 & -3.269 & 320.395 & 22$^m$ & JPL \\ 
\CHtOCHO{} v=0 28(11,17)-27(11,16) A & 346675.64 & -3.269 & 320.396 & 22$^m$ & JPL \\ 
\CHtOCHO{} v=0 28(11,18)-27(11,17) E & 346687.46 & -3.269 & 320.391 & 22$^m$ & JPL \\ 
\CHtOCHO{} v=0 27( 5,22)-26( 5,21) E & 347478.25 & -3.211 & 247.252 & B & JPL \\ 
\CHtOCHO{} v=0 27( 5,22)-26( 5,21) A & 347493.96 & -3.211 & 247.256 & 30 & JPL \\     
\\
g-\CHtCHtOH{} 20( 3,18)-19( 3,17) vt=0-0 & 345648.57 & -3.436 & 242.486 &  27 & JPL \\ 
g-\CHtCHtOH{} 20(10,10)-19(10, 9) vt=1-1 & 346085.56 & -3.549 & 358.843 &  T & JPL \\
g-\CHtCHtOH{} 20(10,11)-19(10,10) vt=1-1 & 346085.56 & -3.549 & 358.843 &  T & JPL \\
g-\CHtCHtOH{} 20( 9,12)-19( 9,11) vt=1-1 & 346183.19 & -3.522 & 335.566 &  T & JPL \\
g-\CHtCHtOH{} 20( 9,11)-19( 9,10) vt=1-1 & 346183.19 & -3.522 & 335.566 &  T & JPL \\
g-\CHtCHtOH{} 20(11,10)-19(11, 9) vt=0-0 & 346383.64 & -3.580 & 379.110 &  5$^m$ & JPL \\ 
g-\CHtCHtOH{} 20(11, 9)-19(11, 8) vt=0-0 & 346383.64 & -3.580 & 379.110 &  5$^m$ & JPL \\ 
g-\CHtCHtOH{} 20(10,11)-19(10,10) vt=0-0 & 346424.58 & -3.548 & 353.452 &  T & JPL \\
g-\CHtCHtOH{} 20(10,10)-19(10, 9) vt=0-0 & 346424.58 & -3.548 & 353.452 &  T & JPL \\
g-\CHtCHtOH{} 20( 9,12)-19( 9,11) vt=0-0 & 346505.34 & -3.521 & 330.252 &  T & JPL \\
g-\CHtCHtOH{} 20( 9,11)-19( 9,10) vt=0-0 & 346505.34 & -3.521 & 330.252 &  T & JPL \\
g-\CHtCHtOH{} 20( 7,14)-19( 7,13) vt=1-1 & 346565.08 & -3.479 & 296.366 & 22$^m$ & JPL \\ 
g-\CHtCHtOH{} 20( 7,13)-19( 7,12) vt=1-1 & 346565.39 & -3.479 & 296.366 & 22$^m$ & JPL \\ 
g-\CHtCHtOH{} 20( 8,13)-19( 8,12) vt=0-0 & 346620.32 & -3.498 & 309.511 &  T & JPL \\
g-\CHtCHtOH{} 20( 8,12)-19( 8,11) vt=0-0 & 346620.33 & -3.498 & 309.511 &  T & JPL \\
g-\CHtCHtOH{} 20( 7,14)-19( 7,13) vt=0-0 & 346816.58 & -3.478 & 291.241 &  12$^m$ & JPL \\ 
g-\CHtCHtOH{} 20( 7,13)-19( 7,12) vt=0-0 & 346816.92 & -3.478 & 291.241 &  12$^m$ & JPL \\ 
t-\CHtCHtOH{} 21( 0,21)-20( 1,20) 	 & 346962.59 & -3.616 & 185.843 &  B & JPL \\ 
g-\CHtCHtOH{} 20( 6,15)-19( 6,14) vt=0-0 & 347147.20 & -3.461 & 275.456 &  13 & JPL \\ 
g-\CHtCHtOH{} 20( 6,14)-19( 6,13) vt=0-0 & 347157.99 & -3.461 & 275.457 &  9 & JPL \\ 
t-\CHtCHtOH{} 14( 3,12)-13( 2,11) 	 & 347350.92 & -3.757 &  99.660 &  B & JPL \\ 
t-\CHtCHtOH{} 21( 1,21)-20( 0,20) 	 & 347445.52 & -3.573 & 185.853 &  41 & JPL \\ 
g-\CHtCHtOH{} 20( 5,16)-19( 5,15) vt=1-1 & 347473.56 & -3.447 & 267.087 &  B & JPL \\ 
\enddata
\tablenotetext{\mbox{}}{$a$: Integrated line intensities (see text for the definition)
measured from the lower disk atmosphere for the reasonably isolated lines detected with
more than 3$\sigma$.  They are the
mean values averaging over a rectangular region (with a size of
\arcsa{0}{2}$\times$\arcsa{0}{05} covering most of the emission) centered at
the lower atmosphere. In this column, the line intensities commented with ``m" are the mean values 
obtained by averaging over 2 or more lines with similar $E_u$ and log $A_{ul}$ for better   
measurements. T:  the lines are tentatively
detected with about 2 $\sigma$ detection and a part of the linear PF
structure. S: the lines are blended with the line(s) of the same molecule at different transitions.
B: the lines are blended with the line(s) of other molecules.
The line intensities here are assumed to have an uncertainty of 40\%.}
\tablenotetext{\mbox{}}{$\dagger$: \tCHtOH{} intensity after removing the small contribution from \CHtDCN{} (see text).}
\end{deluxetable}
\clearpage


\begin{deluxetable}{llll}
\tablecolumns{4}
\tabletypesize{\normalsize}
\tablecaption{Column Densities and Abundances in the Lower Disk Atmosphere
 \label{tab:colabun}}
\tablewidth{0pt}
\tablehead{
\colhead{Species} & \colhead{Excitation Temperature} & \colhead{Column Density} &  \colhead{Abundance$^\dagger$}\\
& (K) & (\cms) & 
}  
\startdata
\CHtOH{}...$^a$  &$160\pm45$ & \scnum{($6.1\pm3.3$)}{17} & \scnum{($1.6\pm0.9$)}{-7}  \\
\CHtOH{}...$^b$  &$150\pm50^c$ & \scnum{($1.3\pm0.7$)}{18} & \scnum{($3.3\pm1.8$)}{-7}  \\
\CHtDOH     &$148\pm19$ & \scnum{($1.6\pm0.7$)}{17} & \scnum{($4.2\pm1.8$)}{-8}  \\
\tCHtOH$^d$ &$150\pm50^c$ & \scnum{($2.5\pm1.4$)}{16} & \scnum{($6.6\pm3.7$)}{-9}  \\ 
\HtCCO      &$150\pm50^c$ & \scnum{($3.8\pm2.0$)}{15} & \scnum{($1.0\pm0.5$)}{-9}  \\
t-HCOOH     &$150\pm50^c$ & \scnum{($2.0\pm1.1$)}{16} & \scnum{($5.3\pm2.9$)}{-9}  \\
\CHtDCN     &$150\pm50^c$ & \scnum{($3.6\pm1.6$)}{14} & \scnum{($9.5\pm4.2$)}{-11} \\
\CHtCHO     &$150\pm50^c$ & \scnum{($1.5\pm0.6$)}{16} & \scnum{($3.9\pm1.6$)}{-9}  \\
\CHtOCHO    &$150\pm50^c$ & \scnum{($3.2\pm1.6$)}{16} & \scnum{($8.4\pm4.2$)}{-9}  \\
\CHtCHtOH   &$150\pm50^c$ & \scnum{($2.7\pm1.2$)}{16} & \scnum{($7.1\pm3.2$)}{-9}  \\
\CHtSH$^e$  &$150\pm50^c$ & \scnum{($9.3\pm5.5$)}{16} & \scnum{($2.4\pm1.4$)}{-8}  \\
\NHtCHO{}$^e$ &$150\pm50^c$ & \scnum{($1.6\pm0.9$)}{15} & \scnum{($4.2\pm2.4$)}{-10}\\
\DtCO{}$^e$ & $150\pm50^c$ & \scnum{($3.0\pm1.6$)}{15} & \scnum{($7.9\pm4.2$)}{-10} \\
\enddata
\tablenotetext{\mbox{}}{$\dagger$: Derived by dividing the column densities of the molecules by
the H$_2$ column density in the disk atmosphere, which is $\sim$ \scnum{3.8}{24} \cms{} (see text).}
\tablenotetext{\mbox{}}{$a$:  Derived from the rotation diagram in Figure \ref{fig:pop}a.
The value derived this way is considered as the lower limit of the \CHtOH{} column density (see text).}
\tablenotetext{\mbox{}}{$b$:  Obtained by multiplying the \tCHtOH{} column density by 50,
which is the $^{12}$C/$^{13}$C ratio obtained in the Orion complex \citep{Kahane2018}.
The value derived this way is adopted for the \CHtOH{} column density (see text).}
\tablenotetext{\mbox{}}{$c$: Assuming an excitation temperature (rotational temperature) of $150\pm50$ K,
which is the mean value derived from the \CHtOH{} and \CHtDOH{} lines.}
\tablenotetext{\mbox{}}{$d$: Estimated after subtracting the small contribution from \CHtDCN{} (see text).}
\tablenotetext{\mbox{}}{$e$: Adopted from \citet{Lee2017COM}, updated for an excitation temperature of $150\pm50$ K.}
\end{deluxetable}

\end{document}